\documentclass[useAMS,usenatbib]{mn2e}
\usepackage{aas_macros}
\usepackage{amssymb,amsmath}
\usepackage{times}
\usepackage[british]{babel}
\usepackage{graphicx}
\roman{enumi}

\begin{document}
\title[X-ray/Ultraviolet/Optical variability of NGC\,4395]{Correlated X-ray/Ultraviolet/Optical variability in the very low mass AGN NGC\,4395}
\author[D. T. Cameron et al.]
{D. T. Cameron,$^1$\thanks{E-mail: dtc1g08@soton.ac.uk} I. M$^{\mathrm{c}}$ Hardy,$^1$ T. Dwelly,$^1$ E. Breedt,$^2$ P. Uttley,$^1$ P. Lira,$^3$ P. Arevalo,$^4$\\
	$^1$ School of Physics and Astronomy, University of Southampton, Southampton, S017 1BJ. UK \\
	$^2$ Department of Physics, University of Warwick, Coventry, CV4 7AL. UK \\
	$^3$ Departmento de Astronom\'{\i}a, Universidad de Chile, Camino del Observatorio 1515, Santiago, Chile \\
	$^4$ Departmento de Ciencias Fisicas, Universidad Andres Bello, Av. Republica 252, Santiago, Chile}

\maketitle
\begin{abstract}
We report the results of a one year {\em{Swift}} X-ray/UV/optical
programme monitoring the dwarf Seyfert nucleus in NGC\,4395 in 2008--2009.  The
UV/optical flux from the nucleus was found to vary dramatically over
the monitoring period, with a similar pattern of variation in each of
the observed UV/optical bands (spanning 1900 -- 5500 \AA).  In particular, the
luminosity of NGC\,4395 in the 1900 \AA\ band changed by more than a
factor of eight over the monitoring period.  The fractional
variability was smaller in the UV/optical bands than that seen in the
X-rays, with the X-ray/optical ratio increasing with increasing
flux. Pseudo-instantaneous flux measurements in the X-ray and each UV/optical band were
well correlated, with cross correlation coefficients of $\ge 0.7$,
significant at 99.9 per cent confidence.  Archival {\em{Swift}} observations from 2006
sample the intra-day X-ray/optical variability on NGC\,4395.  These
archival data show a very strong correlation between the
X-ray and $b$ bands, with a cross-correlation coefficient of 0.84
(significant at $>99$ per cent confidence). The peak in the cross correlation
function is marginally resolved and asymmetric, suggesting that X-rays
lead the $b$ band, but by $\leq$ 1\,hour.  In response to recent (August 2011) very
high X-ray flux levels from NGC\,4395 we triggered {\em{Swift}} ToO
observations, which sample the intra-hour X-ray/UV variability. These
observations indicate, albeit with large uncertainties, a lag of the
1900 \AA\ band behind the X-ray flux of $\sim$400 s.  The tight
correlation between the X-ray and UV/optical lightcurves, together
with the constraints we place on lag time-scale are consistent with the
UV/optical variability of NGC\,4395 being primarily due to
reprocessing of X-ray photons by the accretion disc.
\end{abstract}


\section{Introduction}

Almost since their discovery, active galactic nuclei (AGN) have been
noted to be variable objects, with flux variations seen on a wide
range of time-scales and across the observable electromagnetic
spectrum.  However, the origin of the ultraviolet (UV) and optical
variability in AGN is still a matter of debate. Two mechanisms that
have gained support are i) reprocessing of X-ray emission by the
accretion disc
\citep{Krolik1991,Wanders1997,Collier1998,Sergeev2005,Breedt2009,Breedt2010},
and ii) intrinsic variations of the thermal disc emission
\citep{Arevalo2008,Lira2011}. In scenario i) the X-ray fluctuations drive and
lead the UV/optical variations, however in ii), the perturbations are
produced in the disc itself, for example, by inwardly propagating
accretion rate variations.

In the case of the X-ray reprocessing model, the optical/UV variations
are expected to lag behind the X-ray variations, where the length of
the lag is determined by the light travel time from the central
compact X-ray emission region to the optical/UV emission region in the
disc. Simply from conservation of energy, the absolute luminosity
variation in the optical/UV should be smaller than the luminosity
variation seen in the (driving) X-ray band.  For AGN having
black hole masses in the range $10^{6}$--$10^{8}\,$M$_{\odot}$ (as typically seen in nearby Seyfert galaxies)
the lag time-scales are expected to range from seconds to a few days \citep[e.g.][]{Breedt2009,Breedt2010}. The
X-ray to UV/optical lag time-scales are expected to increase with observed
wavelength \citep[e.g.][]{Cackett2007}, because the longer wavelength
emission is produced at larger radii in the disc.

In the scenario where the UV/optical variations originate in the accretion
disc, then we expect that these inwardly propagating accretion rate
fluctuations will result in the X-ray lagging behind the optical/UV variations. The lag
time-scale is then dictated by the viscous propagation time-scale, which is
always much longer than the light-travel time, typically months or years for
standard accretion discs about black holes having masses $10^{6}$--$10^{8}\:$M$_{\odot}$ \citep{Czerny2006}.

Of course it is likely that both of these processes are occurring
simultaneously within an AGN \citep[e.g.][]{Arevalo2008,Arevalo2009},  but we can hope to test which of these
processes is causing the {\em majority} of the UV/optical variation.  The
nearby Seyfert nucleus in NGC\,4395 is perhaps the ideal laboratory in
which to test these competing hypotheses, because it is dramatically
variable across the X-ray/UV/optical bands, and is thought to contain
a relatively low mass black hole, meaning that the characteristic
time-scales of variations are well matched to the typical durations of
intensive observing campaigns (i.e. $\le 1$ year).

NGC\,4395 is a Seyfert 1.8/1.9 galaxy \citep{Filippenko1989} with a
black hole mass estimated from reverberation mapping, of $(3.6\pm1.1)
\times 10^5 \: $M$_\odot$ \citep{Peterson2005}.  A somewhat smaller black hole
mass is suggested by the apparent lack of a significant galactic bulge in
NGC\,4395, and also from the very short characteristic time-scale
measured in the X-ray band \citep{Vaughan2005}.  The absolute upper
limit on the total mass within the central 3.9 pc, including any
stellar contribution, is $6.6 \times 10^6\: $M$_\odot$, derived from
stellar velocity dispersion measurements \citet{Filippenko2003}.
These black hole mass estimates rank NGC\,4395 as one of the very
lowest mass AGN known. The distance to the galaxy, derived directly
from observations of Cepheid variables, is $4.2\pm0.3$ Mpc
\citep{Thim2004}.  The proximity of NGC\,4395 means that although its
luminosity is low in all wavebands, it is bright enough in the X-ray/UV/optical
to be observed with short {\em{Swift}} snapshots. The very low Galactic extinction,
E($B-V$) = 0.017\,mag, in the direction of NGC\,4395 \citep*{Schlegel1998}
means that it is particularly well placed from the point of view of UV
observations, which will not be heavily attenuated.

NGC\,4395 has long been noted as a particularly variable AGN,
exhibiting large amplitude variations on short time-scales, across many
wavebands. For example, a continuous single orbit (113\,ks) {\em
 XMM-Newton} observation was analysed by \citet{Vaughan2005}. This
observation showed the most extreme X-ray variability ever seen in an
AGN with a fractional root-mean-squared (RMS) amplitude (F$_{var}$) of
more than 100 per cent in the soft X-ray band (0.2--0.7\,keV).  Simultaneous
optical and near-infrared observations are reported by
\citet{Minezaki2006} who estimate the distance to the inner face of
the torus in NGC\,4395 as $\sim$1\,lightday providing an upper limit
to the size of the accretion disc. The broad line region radius
was determined by \citet{Peterson2005,Peterson2006Erratum} to be
$\sim$0.04 pc.

Many of the recent dedicated monitoring campaigns for NGC\,4395 have
measured lightcurves from either a single, or just a few bands spaced
closely in wavelength, and have typically not examined the correlated
broadband variability behaviour.  For instance, the Optical Monitor
imaging data during the long {\em XMM-Newton} observation are not well
suited to a X-ray/UV/optical variability analysis.  Older soft X-ray
observations with {\em ROSAT}, together with a variety of ground based
optical and infrared observations, were presented by
\citet{Lira1999}. They reported large X-ray flux changes between
observations, e.g. by a factor of 2 in 15\,days along with 20 per cent
changes in optical flux from night to night. However these authors
were unable to comment on the correlation between variations in the
optical and X-ray bands due to the non-simultaneity of their data.

The best dataset previously available that probes the short time-scale
relationship between the X-ray and UV/optical emission in NGC\,4395
has been carried out with coordinated {\em Chandra}, Hubble Space
Telescope (HST) and ground based optical observations
(e.g. \citealt{O'neill2006}, \citealt{Desroches2006}), in which the UV
and X-ray datasets overlap by $\sim$10\,ks. After cross-correlating
the 1350 \AA\ and 0.4--8\,keV lightcurves, \citet{O'neill2006} find a
UV lag consistent with zero with an uncertainty of about 25 minutes.
\citet{Desroches2006} present ground based optical $B$ and $V$ band
observations taken in parallel with these {\em Chandra}
observations. The X-ray to optical correlation is rather weak, with a
rather flat cross-correlation function (CCF) having a peak correlation
coefficient of $\sim$0.2. \citet{Desroches2006} estimate that the
optical lightcurves lag the X-rays by $44\pm13$ minutes. The
multi-wavelength results from the {\em Chandra}/{\em HST UV}/optical
campaign are consistent with a scenario where the optical variations
are primarily due to reprocessed X-ray emission. However, these
lightcurves span only a few hours, so these observations clearly do
not fully sample all of NGC\,4395's longterm behaviour. Also, it
should be noted that during these observations NGC\,4395 was rather
faint in the X-ray, with an average flux less than half of that during
the previous $\sim$100\,ks {\em XMM-Newton} observations
\citep{Vaughan2005}. The low flux levels and relatively weak
variability in the X-ray, UV and optical lightcurves reduces the
ability of the {\em Chandra}/{\em HST}/optical dataset to constrain
the physical parameters of the system.

In order to improve our understanding of the X-ray/UV/optical
connection we have carried out a dedicated year-long campaign
monitoring NGC\,4395 using the NASA {\em Swift} observatory
\citep{Swift}.  We present the results of these observations which we
have analysed together with previous archival {\em{Swift}} X-ray and
optical monitoring which intensively samples the intra-day X-ray/optical
variability.  These data are by far the most extensive joint
X-ray/UV/optical observations ever made of NGC\,4395 and, indeed, of
almost any AGN. The multi-band {\em Swift} data allow us to study the
correlations and lags between X-ray and UV/optical bands in
unprecedented detail and on time-scales from a few hundred seconds up to one
year. We note that the {\em{Swift}} X-ray lightcurve for NGC\,4395 has
previously been presented by \citet{NGC4395X-ray}.  In this paper we
concentrate on the cross correlation between the X-ray, UV and optical
wavebands; a fuller analysis of the {\em{Swift}} X-ray dataset will be
presented separately (Dwelly et al. in prep).

The details of the {\em{Swift}} observations are presented in Section
\ref{sec:Observations} and the resulting lightcurves are presented in
Section \ref{sec:Variability}.  In Section \ref{sec:Cross-correlations}
we present the cross-correlation results and we discuss
their implications. In Section \ref{sec:AccDiscMod} we compare the predictions 
of the standard accretion disc model to the observed behaviour.  
 We summarise our results in
Section \ref{sec:Discussion} and present our conclusions.

\section{\em{Swift} Observations}
\label{sec:Observations}
\subsection{The 2008--2009 long term dataset}
We monitored NGC\,4395 with {\em Swift} with $\sim$1\,ks of exposure
time approximately every 2 days during the period 2008 April to 2009 March.
There was a break from 2008 August 6th to 2008 October 21st when
NGC\,4395 was too close to the Sun, thus separating the observations
into two well-sampled sections (referred to hereafter as sections 1
and 2). Useful data were collected by both the X-Ray Telescope
\citep[XRT,][]{XRT} and the Ultraviolet/Optical Telescope
\citep[UVOT,][]{UVOT}.  Each $\sim$1\,ks observation was sub-divided into up to eight
separate `visits', with individual visit lengths
ranging from $\sim$100 s to over 1000 s. In total 254 visits were made.
Within each visit, the UVOT observations were taken through the
$uvw2$, $v$, $u$, and $b$ filters, in that order, with exposure time
ratios of approximately 4:1:1:1 for the $uvw2$, $v$, $u$ and $b$
filters respectively. The majority of the UVOT data were obtained in
`event' mode where the time of arrival and location of each photon is
recorded, meaning that we can post-process these data to make
images/lightcurves in time bins of arbitrary length.

\subsection{The 2006 intensive short time-scale dataset}
We have also analysed a set of intensive {\em{Swift}} observations of NGC\,4395 which were made
from 2006 March 6th to 2006 March 10th, which we refer to hereafter as the
`2006' observations.  Throughout this 5 day monitoring period one
{\em{Swift}} visit of $\sim$1\,ks was made per orbit (i.e. once every
$\sim$96\,minutes), totalling 61 visits.  All of the UVOT observations
of NGC\,4395 during this period were made through the $b$ filter in
`image' mode, in which a single UVOT integration was obtained per visit.

\subsection{The 2011 target of opportunity dataset}
Very recently (August 2011) we have triggered a set of {\em Swift} ToO observations
of NGC\,4395 in response to a high X-ray flux state. These data
consist of 6 visits spanning $\sim$13\,hours, and most importantly,
include two relatively long visits of length $\sim$2500 s. The UVOT was
operated with the $uvw2$ filter and in event mode during these ToO
observations.

\subsection{Data Reduction Procedure}
The UVOT raw data were cleaned, flat-fielded and corrected
for mod-8 (fixed pattern noise due to sub-pixel centralisation)
noise using the standard {\em{Swift}} UVOT
pipelines\footnote{http://swift.gsfc.nasa.gov/docs/swift/analysis/}.
From an initial examination of the UVOT images, we noticed that a
higher than expected fraction of the visits were affected by poor
pointing stability caused by loss of guide star lock, probably due to
the scarcity of bright stars at the high Galactic latitude of NGC\,4395.
Fortunately, the majority of the affected observations were made in
`event mode' allowing the spacecraft attitude information to be
corrected on 10--20 second time-scales by tracking the apparent
locations of bright stars within the UVOT field of view.  For each
{\em Swift} observation a series of short snapshot images (10 seconds
for UVOT observations in the $u$, $b$ and $v$ filters, and 20 seconds
for the $uvw2$ filter) were created from the raw UVOT event
lists. Each snapshot was searched for pointlike sources, the apparent
positions of which were cross-correlated with a list of bright
($u'_{AB} < 18$) reference stars (obtained from the SDSS catalogue
server\footnote{http://cas.sdss.org/astrodr7/en/}).  A linear offset
(in RA and Dec) to the nominal pointing of the snapshot was then
calculated from the crossmatches. A new spacecraft attitude file was
then created for each {\em Swift} observation by applying a filtered
and interpolated version of the snapshot offsets to the original
spacecraft attitude file.  Our astrometric correction method was
adapted from routines developed by S. R. Oates for the Swift Gamma Ray
Burst catalogue \citep{Roming2009}.  This astrometric correction process was seen to
improve the registration of UVOT images that were not
affected by guide star lock, and so for completeness was applied uniformly to the
entire UVOT dataset.  For those visits where the UVOT data were
obtained in `image' mode, we calculated only a single pointing
correction possible per UVOT exposure.  We discarded the data from the few
visits where the UVOT was in image mode which also showed signs of star
trailing.

The {\em Swift}-XRT data reduction procedure will be fully described
in Dwelly et al. (in prep).  A brief description of our method can be
found in Section 2 of \citet{Fabian2011}. We note that throughout this
work we use X-ray lightcurves that have been corrected for background,
vignetting, aperture losses and for the effects of bad pixels.

\section{Ultraviolet and optical variability}
\label{sec:Variability}
Aperture photometry was performed using the {\em{Swift}} UVOT tool
{\sc UVOTSOURCE}\footnote{http://heasarc.nasa.gov/lheasoft/ftools/headas/uvotsource.html},
with a 3 arcsec radius aperture.
The background region was chosen to be representative of the sky
near the galaxy.
Previous {\em HST}-WFPC2 (F815W, $\sim$ $I$ band) observations have
revealed that in addition to the central point source, there is a
significant galactic contribution within the innermost few arcseconds
of NGC\,4395 \citep{Filippenko2003} which is expected to contaminate
our UVOT aperture photometry of the nucleus. Note that the point
spread function (PSF) of UVOT has a FWHM of $\sim$2.5\,arcsec. We have
therefore used archival {\em HST} observations in the F330W filter
(close to UVOT $u$ band), the F439W filter (close to UVOT $b$ band)
and the F555W filter (close to UVOT $v$ band), in order to estimate
the constant contribution from non-nuclear light to our UVOT aperture
fluxes. This is possible as the {\em HST} resolution is $\sim50\times$
better than {\em{UVOT}}.  We downloaded processed images from the {\em
  HST} Legacy Archive
website\footnote{http://hla.stsci.edu/hlaview.html}.  We used the
{\textsc{GALFIT}} software \citep{Galfit1,Galfit2} to fit a galaxy
profile model to the central regions (an 18 arcsec by 15 arcsec rectangle) of
NGC\,4395, adopting the three profile model components of
\citet{Filippenko2003}, namely a nuclear point source, a Sersic
component representing the nuclear stellar cluster, and an exponential
disc. Bright irregular features and Galactic stars were masked out
during the fitting process. The contribution from the nuclear point
source component was subtracted from the image, which was then
convolved with the appropriate UVOT PSF model taken from the
calibration database.  Finally, the non-nuclear flux within the
3\,arcsec radius science aperture was calculated from the convolved
image, converting to physical units using the standard {\em HST}
zeropoints.  We measure the following non-nuclear flux density
contributions within the UVOT science aperture: $4.9 \times
10^{-16}$\,erg\,cm$^{-2}$\,s$^{-1}$\,\AA$^{-1}$ in the $u$ band, $4.9
\times 10^{-16}$\,erg\,cm$^{-2}$\,s$^{-1}$\,\AA$^{-1}$ in the $b$ band
and $5.0 \times 10^{-16}$\,erg\,cm$^{-2}$\,s$^{-1}$\,\AA$^{-1}$ in the
$v$ band. No suitable high resolution UV images were available, so to
estimate the non-nuclear flux contamination in the $uvw2$ band we
normalised the Scd galaxy template of \citet{StarlightGal} to the
non-nuclear contribution in the $b$ band, giving an expected
non-nuclear flux of $4.8 \times
10^{-16}$\,erg\,cm$^{-2}$\,s$^{-1}$\,\AA$^{-1}$ in the $uvw2$ band.
We subtract these non-nuclear flux estimates from the aperture fluxes
calculated within {\textsc{UVOTSOURCE}}. We expect this procedure to
be accurate to around
$0.1\times10^{-16}$\,erg\,cm$^{-2}$\,s$^{-1}$\,\AA$^{-1}$ based on the difference between
results produced for subtractions using the 2 archived {\em{HST}} images with the B-filter.
Any systematic offset imprinted on the UVOT data by this method will not effect
the cross-correlation analysis performed in Section \ref{sec:Cross-correlations}.
However, any residual systematic flux offsets may effect the results of the accretion disc modelling
discussed in Section \ref{sec:AccDiscMod}.

In order to verify the stability of the UVOT photometry measured by
{\textsc{UVOTSOURCE}} and to refine the default aperture corrections,
we have examined the lightcurves of a relatively isolated field star
that has similar apparent magnitude to the nucleus of NGC\,4395.  We extracted
$uvw2$, $u$, $b$ and $v$ band lightcurves for this star from the
2008--2009 UVOT dataset using 3 arcsecond apertures and an annular
background region, ignored the few UVOT images where the star falls off the edge of the detector. 
We found that the UVOT photometry for this star is
stable, with no obvious long-term drifts. We show a summary of the
properties of the measured photometry in Table \ref{tab:teststars}.
Note that the scatter of the measured fluxes in each of the $uvw2$,
$u$ and $v$ filters are comparable to the the nominal uncertainties
estimated by the {\textsc{UVOTSOURCE}} tool.  In the $b$ filter
the error estimates are on average 1.72 times smaller than the true
photometric scatter, suggesting that the errors are underestimated for
this filter.  This additional error is included when the fluxes are corrected for the aperture size.  We also
measured the lightcurves of the star using the standard 5\,arcsec
radius aperture, for which the UVOT counts to flux density relation
has been very accurately calibrated
\citep{Poole2008,Breeveld2010}. Using these values we measure aperture
correction values for the 3\,arcsec radius aperture. We find that the
fraction of light in the 3\,arcsec radius aperture is marginally
smaller for the $uvw2$ filter than given by the standard {\em Swift}
calibrated value (1.13, 1.10, 1.11, 1.09, in the  $uvw2$, $u$, $b$ and $v$ filters
 respectively, \citealt{Poole2008}). However, we find that the aperture corrections in
the $u$, $b$ and $v$ filters agree closely with those given by the
standard calibration.  We use the aperture correction values
calculated for the test star to correct the 3\,arcsec radius aperture
photometry for NGC\,4395.

\begin{table}
 \caption{Tests of the accuracy of our UVOT photometry as determined
   from measurements of a star in the field of NGC\,4395.  $N$ is the
   number of separate exposures considered for each filter,
   $\overline{f}$ is the mean flux of the star in each filter in units
   of $10^{-15}$\,erg\,s$^{-1}$\,cm$^{-2}$\,\AA$^{-1}$.
   $\overline{|f_i-\overline{f}|/\sigma_i}$ is the mean of the ratio
   of the measured photometric scatter to the nominal uncertainties
   reported by the {\textsc{UVOTSOURCE}} tool.  The $\overline{cf}$
   are the mean aperture correction factors we measure for the star
   for a 3\,arcsec radius aperture.}
 \label{tab:teststars}
 \begin{tabular}{@{}lcccc}
  \hline
Filter   &  $uvw2$ & $u$ & $b$ & $v$\\
  \hline
$N$              &     197     &    155       &     110     &    153    \\
$\overline{f}$              &     1.72    &    6.43      &     8.72    &    7.53   \\
$\overline{|f_i-\overline{f}|/\sigma_i}$ & 0.815 & 0.94 & 1.72 & 1.08 \\
$\overline{cf}$      &     1.17$\pm$0.03   &    1.10$\pm$0.01      &     1.10$\pm$0.01    &    1.10$\pm$0.01   \\
  \hline
 \end{tabular}
\end{table}

The UVOT fluxes were
corrected for the Galactic reddening of E($B-V$) = 0.017 mag, by applying the formulae
from \citet{Roming2009}, to calculate the appropriate correction in each UVOT filter.
The fully corrected UVOT
lightcurves are presented in Fig. \ref{fig:longlc} together with the
2--10\,keV XRT lightcurve.

\subsection{Long-term variability}
\label{sec:longterm_data}
\begin{figure*}
 \includegraphics[width=180mm,angle=0]{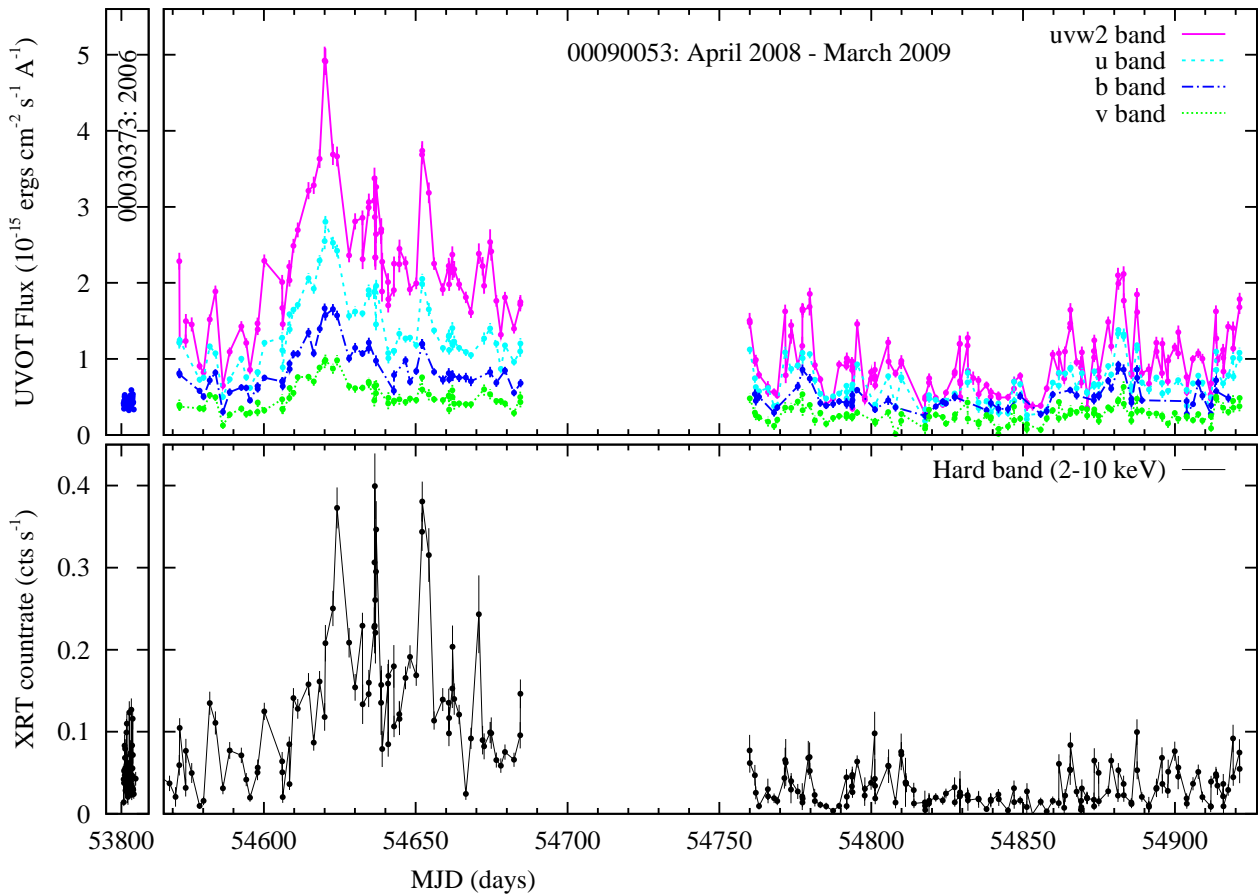}
 \caption{{\em Top panel:} Light curves of the nucleus of NGC\,4395 for the $uvw2$, $u$, $b$ and $v$ UVOT bands.
 {\em Bottom panel:} Simultaneous 2--10\,keV X-ray lightcurve of NGC\,4395, with countrates normalised
to account for aperture losses, bad pixels and vignetting.
A single datapoint is shown for each individual {\em Swift} visit.}
 \label{fig:longlc}
\end{figure*}

During the 2008--2009 observation period, NGC\,4395 showed significant
variability across all UVOT filters, Fig. \ref{fig:longlc}.  We
calculated the fractional root mean square amplitude (F$_{var}$)
separately for each of the two contiguous sections of the lightcurve,
and the results are shown in Table \ref{RMS}. The fractional
variability for each filter and during both sections are consistent.
This is expected for observations of similar lengths \citep{fRMS}.
The peak-to-trough variation during section 1 is larger than in
section 2, with the $uvw2$ band peak flux $\sim8\times$ greater than
the minimum flux.  During section 1 NGC\,4395 reached a UV/optical
flux approximately twice that of the typical peak fluxes of previous
observations \citep{Lira1999,O'neill2006}.

\begin{table}
 \caption{Central wavelength and fractional RMS variability for the lightcurves measured in each {\em{Swift}}
 UVOT filter.  $ \lambda $ is the central wavelength of the filter.
 F$_{var1}$ and F$_{var2}$ are the fractional root mean square
 variability for the 1st and 2nd sections of the 2008--2009 observations
 respectively.}
 \label{RMS}
 \begin{tabular}{@{}lccc}
  \hline
  Filter & $\lambda$ (\AA) & Fvar$_1$ & Fvar$_2$ \\
  \hline
  $uvw2$ & 1928 & $0.367\pm0.017$ & $0.427\pm0.032$ \\
  $u$ & 3465 &  $0.355\pm0.017$ & $0.395\pm0.032$ \\
  $b$ & 4392 & $0.347\pm0.022$ & $0.325\pm0.035$ \\
  $v$ & 5468 & $0.347\pm0.042$ & $0.420\pm0.089$ \\
  \hline
 \end{tabular}
\end{table}

\subsection{Short-term variability}
\begin{figure}
 \includegraphics[width=85mm,angle=0]{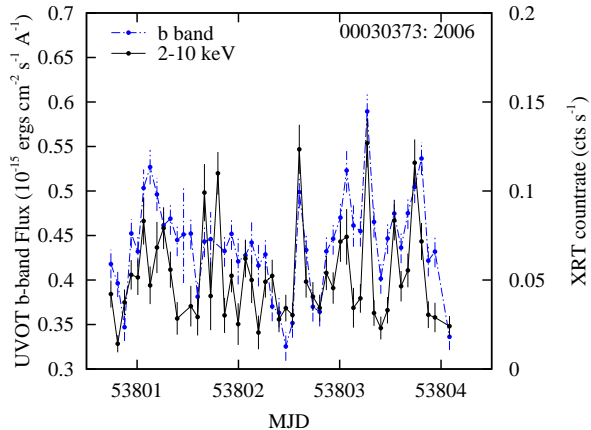}
\caption{The UVOT $b$ band, and XRT 2--10\,keV light curves of
NGC\,4395 from the 2006 intensive monitoring data.  One point is shown
per {\em Swift} visit, which are typically separated by approximately
96 minutes.  Note that the $b$ band lightcurve is shown with a
non-zero intercept (left hand scale) to emphasise the variations.}
 \label{fig:shortlc}
\end{figure}

We reduced the {\em Swift} UVOT and XRT data obtained during the intensive
monitoring observations in 2006 in exactly the same way as
described in Section \ref{sec:longterm_data}. The 
$b$ filter and X-ray lightcurves are shown in Fig. \ref{fig:shortlc}.  The fractional RMS of the $b$
band lightcurve during this period was only $0.122\pm0.006$, which is 
less than a third of variability observed during the 2008--2009
observations.  This reduction is to be expected due to the red noise
nature of AGN lightcurves, in which the largest amplitude of variability
is found on the longest times-cales. However, significant variability
with an amplitude greater than the measurement error is still seen on
a time-scale of one or two orbits within the 2006 dataset.

\subsection{Very short-term variability}
For those UVOT data that were collected in `event' mode we are able to
generate lightcurves with arbitrary time binning.  In principle, this
permits us to investigate lags on time-scales shorter than the length
of a whole {\em Swift} visit (typically $\sim$100--1000\,s), but, due
to the typically rather low XRT countrate measured for NGC\,4395 one
cannot usually sub-divide the {\em Swift} visits whilst retaining a
high signal-to-noise lightcurve.  However, during the brightest
outbursts from NGC\,4395 we measure countrates $>0.5$\,cts\,s$^{-1}$
in the XRT, and so lightcurves with binning of 200\,s are feasible.  We
first examined the short time-scale X-ray/UV/optical variability in the
2008--2009 dataset during the epochs when NGC\,4395 was at its
brightest. Unfortunately, because the exposure time in each {\em
  Swift} visit was spread between four UVOT filters the range of lag
time-scales covered was small. It was not possible to derive a useful
measure of the the lag between any single UVOT waveband and the X-ray
band.  There was a very small bump in the cross-correlation hinting that the lag could be $\sim 200$\,s.

\begin{figure}
\includegraphics[width=85mm,angle=0]{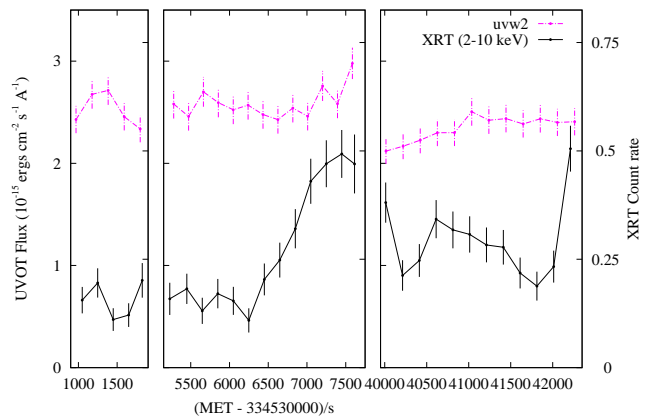}
\caption{The UVOT $uvw2$ band, and XRT 0.5--10\,keV light curves of
NGC\,4395 from the 2011 August ToO observations.  One point is shown
per 200s.}
 \label{fig:veryshortlc}
\end{figure}

To improve on these tentative findings we instigated a new long term
{\em Swift} monitoring programme, (starting April 2011 and scheduled
to finish in March 2012) to search for (and respond to) occasions when
NGC\,4395 was sufficiently bright that X-ray/UV variability on
100--200\,s time-scales could be detected with high confidence.  Our
monitoring revealed that NGC\,4395 had again reached a high X-ray flux
level in mid-August 2011 and so we triggered a pre-approved ToO to
obtain long ($>2$\,ks) continuous observations using just the $uvw2$
filter. Unfortunately NGC\,4395 was just about to enter the {\em
  Swift} Sun avoidance region, and so we only obtained $\sim7$\,ks of
data. The resulting lightcurves are shown in
Fig. \ref{fig:veryshortlc}.

\subsection{X-ray flux - UV/optical flux relation}
\begin{figure*}
 \includegraphics[height=180mm,angle=270]{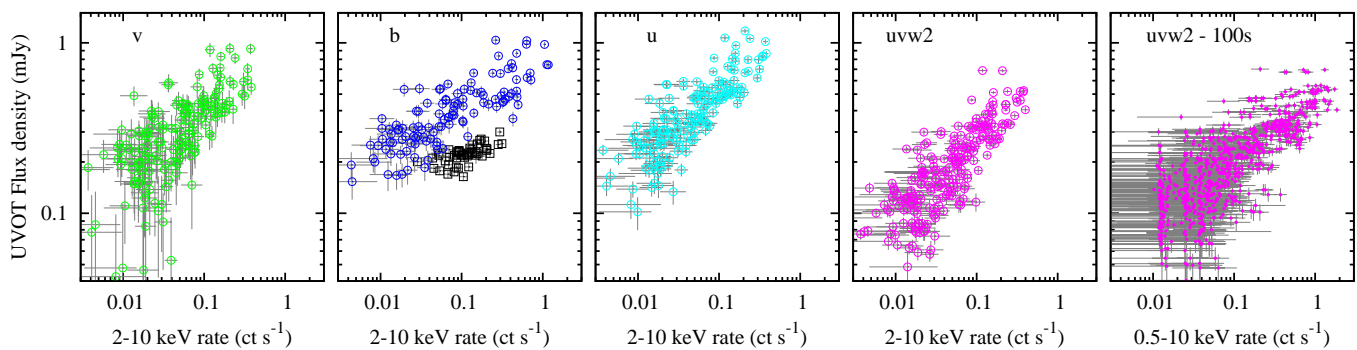}
 \caption{Plots of the 2--10\,keV X-ray countrate versus the instantaneous
 UV/optical flux density. We show data from the long time-scale monitoring campaign
 (2008--2009), with the UVOT filter indicated in the
 top left corner of each panel. In the first four panels from the left
 we show one data point per {\em Swift} visit.  Note that in the $b$ band panel we 
 also show data from the short time-scale monitoring observations in
 2006 (black squares). In the rightmost panel we show the
 X-ray and $uvw2$ data sampled on 100s time-scales. In that rightmost panel we
 plot the full band 0.5--10\,keV count rate to maximise the number of photons per 100s bin
 The grey lines represent the 1$\sigma$ uncertainties.}
 \label{fig:fluxflux}
\end{figure*}

In Fig. \ref{fig:fluxflux} we show the relationships between the quasi-simultaneous
X-ray countrate and UVOT fluxes for the nucleus of NGC\,4395.
We can see there is a tight correlation between the X-ray and UV/optical bands 
with very few outlying points.

In Fig. \ref{fig:fluxflux} we also show the X-ray versus $b$ band
relationship measured during the 2006 intensive monitoring campaign.
The 2006 data clearly follow a different track to the 2008--2009
measurements, with a much lower $b$-band flux for a given X-ray
countrate, with an apparently flatter slope. As a first step in trying
to understand the difference we parametrise the relationship by
$F_{b}=A +B\, F_{X}^{C}$. Here $A$ could be any constant component in
the $b$-band flux. The fitted value of $A$ is similar for both
datasets ($\sim0.15\pm0.06$ mJy) and the exponents ($C$) are also
similar ($\sim0.6$), but the slopes ($B$) differ ($0.4\pm0.3$\,mJy\,(ct\,s$^{-1}$)$^{-1}$
for the 2006 data and $1.2\pm0.15$\,mJy\,(ct\,s$^{-1}$)$^{-1}$ for the 2008--2009 data).
We note that due to the large spread in the data the fits are statistically poor and so
we do not place too much significance on these values.

Although the exact shape of the $b$-band/X-ray relationship is not
well defined, the lower $b$-band flux for a given X-ray countrate in the
2006 observations is clear.  This could be due to there being a reduced
long-term average accretion rate during these observations relative to that in the
2008--2009 observations.  This would reduce the temperature of the accretion
disc and thus the flux in the $b$-band from the disc, even if the X-ray countrate is similar.

\section{Cross-correlations}
\label{sec:Cross-correlations}
We have examined in detail the cross correlation between the
X-ray and UV/optical lightcurves.  We calculate the cross-correlation
function (CCF) using the hard X-ray (2--10\,keV) count rate as a proxy
for the intrinsic X-ray luminosity. The hard band countrate is
expected to be a better representation of the intrinsic luminosity
because it is less sensitive to obscuration (Dwelly et al, in prep).
We used the discrete
cross-correlation method \citep[DCF,][]{Edelson1988} adopting 2 day
binning for the 2008--2009 long-term monitoring data, and 96 minute
binning (similar to the {\em Swift} orbital period) for the 2006
intensive monitoring data. We have examined the range
-40$<$lag$<$+40\,days for the long term monitoring dataset, and
-1$<$lag$<$+1\,day for the intensive monitoring.  We have also
examined shorter time-scales by subdividing event mode data, see
Section \ref{sec:veryshort} below.

\subsection{Long times-cales}
\subsubsection{X-ray - UV/optical cross correlation functions}
\label{sec:zerolag_ccfs}
\begin{figure*}
 \includegraphics[width=180mm,angle=0]{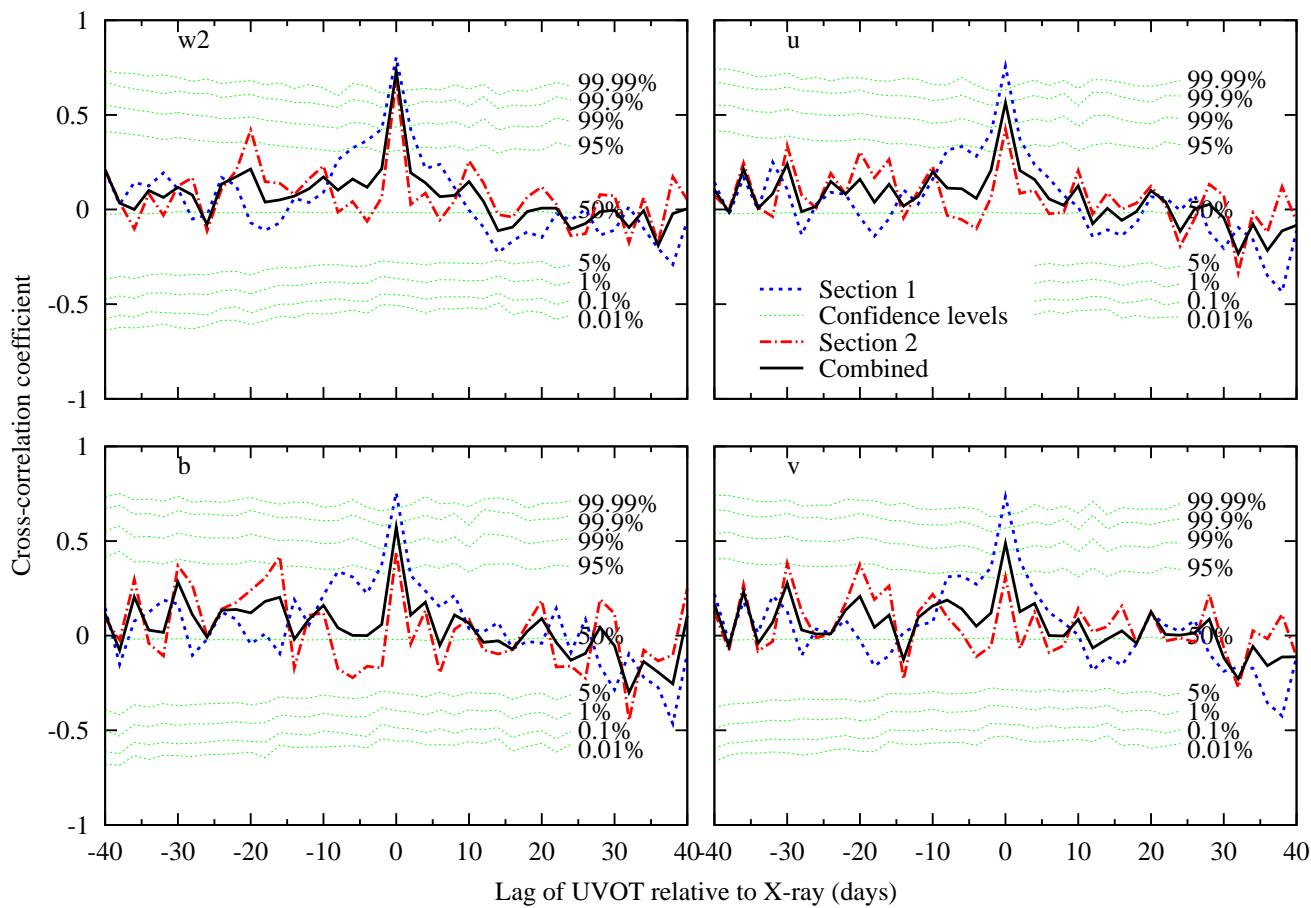}
 \caption{ The long-term cross-correlation functions between the
2--10\,keV X-ray lightcurve and the lightcurves in the $uvw2$, $u$, $b$ and $v$ UVOT bands. 
The UVOT band is indicated in the top left of each
panel. Separate curves are plotted for the the first (blue dashed
line) and second (red dot-dashed line) sections of the 2008--2008
dataset, as well as for the combined CCF (black solid line).  The
dotted (green) contours are derived from Monte Carlo simulations, and
indicate the cross correlation values that a CCF feature must have to
be considered significant at the given confidence level. The contours
shown are appropriate for a single trial (i.e. at a given lag), and are for the
first section of the 2008--2009 observations.}
 \label{fig:ccflong}
\end{figure*}

In Fig. \ref{fig:ccflong} we show the cross correlation functions
calculated by correlating the 2--10\,keV X-ray 2008--2009 lightcurve
with each of the UVOT $uvw2$, $u$, $b$ and $v$ band lightcurves.  We
have first calculated a separate CCF for each of the two contiguous
sections of the 2008--2009 {\em Swift} dataset.  The combined CCF for
the entire 2008--2009 dataset was determined by taking the average
value of the CCFs for each of the two sections, weighted by the
relative time spanned by each section.

We see that each of the CCFs (in each UVOT filter and for each
subsection of the data) have a similar structure, and all show a very
strong sharp peak at zero lag. The temporal resolution of our
2008--2009 {\em Swift} lightcurves is of order 2 days, so the peaks in
the CCFs are constrained to be less than 2 days, i.e. consistent with zero lag.
These small lags are consistent with a reprocessing scenario as long as the
light travel time from the central X-ray source to the UV/optical emission region
is less than 1 day.

For a standard thin accretion disc with a constant accretion rate \citet{Shakura1973} 
predict the disc temperature, in Kelvin,
 as a simple function of radius.
\begin{equation}
\label{eq:T(R)}
T(R) = 
1.2\times10^8 
\left[\frac{\dot{m}}{\dot{m}_{Edd}} \frac{M_\odot}{M_{BH}} \left(\frac{R_g}{R}\right)^3 \left(1-\sqrt{\frac{6R_g}{R}}\right)\right]^{1/4}
\end{equation}
Where $M_{BH}$ is the black hole mass, $M_\odot$ is the mass of the sun, $\dot{m}$ is the accretion rate and $\dot{m}_{Edd}$ is the Eddington accretion rate. \\
So, for a BH with mass $10^5$\,M$_\odot$ having an accretion rate of 1 per cent Eddington, we expect
from the standard thin disc model that the part of the disc
which is emitting at $\lambda_{\mathrm{peak}}$ will lie at a radius,
\begin{equation}
R(\lambda_{\mathrm{peak}})=625 R_g\left(\frac{M_{BH}}{10^5M_{\odot}}\right)^{-\frac{1}{4}}\left(\frac{\dot{m}_{Edd}}{0.01}\right)^{\frac{1}{4}} \left(\frac{\lambda_{\mathrm{peak}}}{3500\mathrm{\AA}}\right)^{3/4}
\label{eq:D}
\end{equation}
from the black hole. Note that we have neglected the
$\sqrt{\frac{6R_g}{R}}$ term as it is $\ll 1$ at the radii that are
responsible for the bulk of the UV/optical emission.  Therefore, for
NGC\,4395 with an assumed BH mass of $10^5$\,M$_\odot$, we expect that
the part of the disc that is emitting at a peak wavelength of
3500\AA\ (i.e. approximately the UVOT $u$ band) will lie at a radius
of 625$R_g$, which is is around 300 light seconds.
Therefore the constraints on the lag time-scale derived from the 2008--2009 dataset ($\leq 2.0$\,days) 
are fully consistent with the expected reprocessing time-scale.

Close examination of the CCF peaks shown in Fig. \ref{fig:ccflong}
reveals that there is a slightly asymmetric shape to the CCF peaks
derived from the first section of the 2008--2009 dataset. This
asymmetry is apparent in the CCFs for all four UVOT bands, appearing
in the sense that the CCF peak declines more slowly towards UV/optical
leads than towards X-ray leads.  This asymmetry is likely caused by
the section of the lightcurves near MJD $\sim$54600 (see
Fig. \ref{fig:longlc}) where the UV/optical lightcurves start to rise
earlier (at MJD $\sim$54595) than the X-ray lightcurve rise (at MJD
$\sim$54608).  We do not see the asymmetry in the CCF peak 
in the second section of the 2008--2009 dataset.

\begin{table}
\caption{The strength of the X-ray to UV/optical cross correlation at
  zero lag, as derived from the 2008--2009 dataset.  DCF is the discrete 
  cross-correlation coefficient at zero lag
  and Sig is its local significance. Results are shown separately for each section of the longterm {\em Swift} 
  monitoring (section 1 is April
  2008 -- August 2008 and section 2 is October 2008 -- March 2009). The significances are calculated from
  the percentage of 100,000 Monte-Carlo simulated lightcurves having
  smaller cross-correlation values at zero lag than the observed DCF
  values. }
 \label{ccfstrength}
 \begin{tabular}{@{}lcccc}
  \hline
              & \multicolumn{2}{c}{Section 1} & \multicolumn{2}{c}{Section 2} \\
  UVOT filter & DCF & Sig & DCF & Sig \\
  \hline
  $uvw2$ & 0.80 & $99.9$ per cent & 0.68 & $99.9$ per cent \\
  $u$ & 0.76 & $99.9$ per cent & 0.42 & $99.9$ per cent \\
  $b$ & 0.76 & $99.9$ per cent & 0.44 & $99.8$ per cent \\
  $v$ & 0.73 & $99.9$ per cent & 0.31 & $98.4$ per cent \\
  \hline
 \end{tabular}
\end{table}

We see from Table \ref{ccfstrength} that there is a trend of
decreasing peak CCF strength with increasing central wavelength of the
UVOT filter.  This trend is consistent with the reprocessing model
since the longer wavelength light is believed to come from the cooler
regions of the disc further away from the central X-ray emission
region.  Increasing the distance of the reprocessor, from the black hole,
results in the reprocessing region subtending a smaller solid angle as
seen from the X-ray source, reducing the fraction of the X-ray emission
that can be reprocessed.  For reprocessing of X-ray photons from a point
source by a disc, lags are observed that are dependent on the increased
travel time to the reprocessing material and also on the distance from
the reprocessing material to the observer.  For example, the predicted
lag for a disc inclined at an angle $\theta$ where $\phi = 0$ is the
most direct line to the observer is given by
\begin{equation}
 \Delta t = \frac{\sqrt{H_x^2+R^2} + H_x \cos \theta - R \sin \theta \cos \phi}{c}
\end{equation}
Where $\Delta t$ is the increased travel time for the reprocessed emission compared
with the direct X-ray emission, H$_x$ is the height above the disc, R is the radius
of the reprocessor, $\theta$ is the inclination angle, $\phi$ is the azimuthal angle
and $c$ is the speed of light.
The longer wavelength emission arises from larger radii in the disc increasing the
spread of light travel times for a disc with non-zero inclination.
The optical emission region also spans a larger range of radii
than the UV emission region.  The combination of these factors
results in the transfer function relevant to the reprocessing of a
$\delta$-function X-ray impulse into UV/optical emission being broader
for longer wavelengths.

We have assessed the significances of the observed CCF peaks using
Monte-Carlo simulations \citep[see e.g.][]{Breedt2010}. We generate
simulated X-ray lightcurves according to the method of
\citet{Timmer1995}, adopting the X-ray power spectral density
parameters for NGC\,4395 determined by \citet{Vaughan2005}, scaled to
the mean and RMS of the observed 2--10\,keV {\em Swift}-XRT
lightcurve, and taking account of the actual sampling pattern in the
{\em Swift} dataset. The simulated X-ray lightcurve was then cross
correlated with each of the observed UVOT lightcurves to produce a
simulated CCF, using the same CCF time axis binning as adopted for the
actual science data.  This process was repeated for a large number
($>10^5$) of simulated X-ray light curves. The resultant ensemble of
simulated CCFs, which by construction should contain no real
correlated signal, allows us to calculate the probability of finding a
particular correlation coefficient at a given lag time by chance.  A
separate set of simulations is carried out for each of the
sub-sections of the data (i.e. the two sections of the 2008--2009
monitoring and the 2006 dataset are all treated separately).
If we decide in advance which lag we are searching for, e.g. zero lag,
then we can estimate the significance of the zero-lag peaks by
determining the fraction of simulated lightcurves that have a CCF
value (at zero lag) that exceeds the measured peak CCF value.
The CCF strength, and the statistical significances of the observed zero-lag peaks are shown in
Table \ref{ccfstrength}.  We find that for the first section of the
2008--2009 dataset, and for each UVOT filter, the detected CCF peaks at zero lag
each have a significance of at least 99.9 per cent.
During the second section of the 2008--2009 dataset the peak cross-correlation strength is
somewhat reduced for all four UVOT filters (see Table \ref{ccfstrength}).
However, the peak at zero lag still has a significance of at least
99 per cent for all but the $v$ band, which itself is significant at $>98$ per cent confidence.

\subsubsection{UVOT--UVOT cross-correlation functions}
\begin{figure}
 \includegraphics[width=85mm,angle=0]{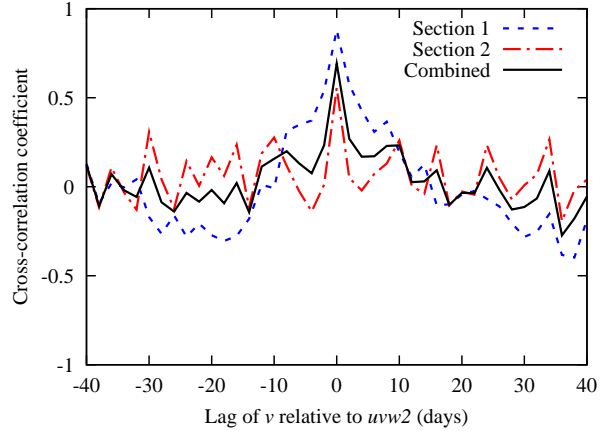}
 \caption{The cross-correlation function between the lightcurves in the $uvw2$ and $v$
 filters, derived from the 2008--2009 UVOT monitoring of NGC\,4395. Note the very
 strong cross-correlation at zero lag. Positive lags would indicate that $v$ lags behind $uvw2$. Section 1 and Section 2
 refer to the first and second contiguous sections of the 2008--2009 dataset respectively.}
 \label{fig:UVOTccf}
\end{figure}

We have also used the 2008--2009 lightcurves to calculate the cross correlations {\em between} the UVOT bands.
An example UV/optical CCF (between the $v$ and $uvw2$ bands) is shown in Fig. \ref{fig:UVOTccf}.
We find that all UVOT bands show very strong correlations with each other
and the CCFs calculated for any combination of UVOT bands show a peak with 
a lag of zero days, consistent with the hypothesis that the
majority of the optical emission is produced by reprocessing of the
X-ray emission.  If the disc fluctuations were the dominant cause of the UV/optical
variability the longer wavelength emission would be expected to lead the shorter
wavelength emission with lags dependent on the viscous time-scales for an accretion
disc.  These are substantially longer than the observed lags. 
We see in Fig. \ref{fig:UVOTccf} that the peak of the CCF is stronger and broader 
when calculated for just the first section of the 2008--2009 dataset (this is also true for most combinations of UVOT bands).
We do not see any indication that there is a strong asymmetry of the main CCF peak, but 
weaker features do appear at lags of $\pm$18--20 days in the CCF calculated for the second section of the 2008--2009 dataset.

\subsection{Short time-scale cross-correlation functions}
\begin{figure}
 \includegraphics[width=85mm,angle=0]{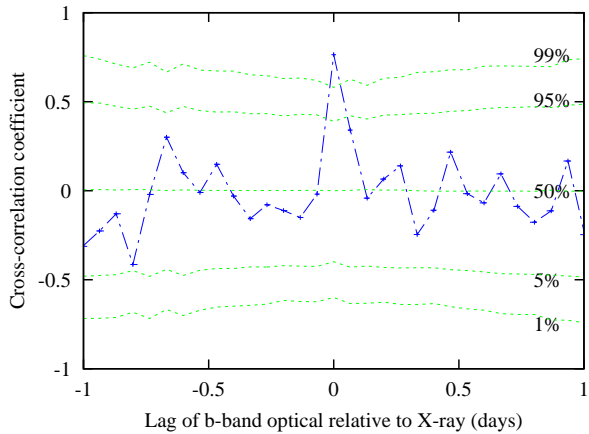}
 \caption{The short time-scale ($<1$\,day) cross-correlation function
 between the X-ray and $b$ bands calculated for the 2006 intensive
 monitoring dataset (blue dot-dashed curve).  Significance levels
 derived from Monte Carlo simulations are shown with (green) dotted
 lines.  The CCF clearly shows a significant and very narrow peak at
 zero lag. The points are binned on 96 minute time-scales, which
 corresponds to approximately one {\em Swift} orbit.}
 \label{fig:ccfearly}
\end{figure}

We have examined the intra-day relationship between the optical and X-ray emission from NGC\,4395 using the
2006 {\em Swift} intensive monitoring dataset.
We calculated the cross-correlation function between the 2--10\,keV X-ray
and the UVOT $b$ band lightcurves using the DCF method, with a bin size of 96 minutes (equal to the orbital period of the
{\em{Swift}} satellite). We plot the result in 
Fig. \ref{fig:ccfearly} and see that the CCF is very strongly peaked at
zero lag, with a peak cross-correlation coefficient of 0.84, dropping rapidly either side of zero lag.
This sharp drop in the CCF indicates that the peak
Therefore we infer that peak cross-correlation between the $b$ band and X-ray lightcurves
must occur at a lag shorter than $<48$min (i.e. half the bin time). We note that for a $\sim10^5$ M$_\odot$
black hole,  48 light minutes corresponds to $6000$\,R$_g$.
The statistical significance of the zero lag peak in the CCF is $>$99 per cent,
determined using a set of Monte Carlo simulations (similar to those described in Section \ref{sec:zerolag_ccfs}).

Close examination of Fig. \ref{fig:ccfearly} reveals that the peak is slightly asymmetric 
towards a positive lag, hinting that
the optical lightcurve lags the X-ray lightcurve, consistent with the results of
\citet{Desroches2006}.  The shape of the CCF peak derived from the
2006 {\em Swift} dataset is sharper and reaches a much higher correlation
coefficient than the CCF presented in Fig. 13 of \citet[][who used
X-ray and UV data from {\em Chandra} and {\em HST} respectively]{Desroches2006}. The 2006 {\em Swift} CCF drops
rapidly to near zero for lags with $|t_{lag}|>2$ {\em Swift} orbits, and is well determined out to lags of $\pm
1$\,day. Therefore, the X-ray/optical CCF derived from the 2006 {\em Swift} dataset constrains very strongly the location of the
reprocessing region relative to the central X-ray source.

The disc reprocessing transfer function relating the X-ray input
to the UV/optical emission can be parametrised fairly accurately by
a very sharp rise followed by an exponential decay
\citep{Breedt2009}. To estimate the width of this function we convolved
the hard (2--10 keV) X-ray lightcurve from 2006, with exponential functions with various decay time-scales
and cross-correlated the resulting lightcurves with the observed
$b$-band lightcurve. We find that the peak value of the
cross-correlation function rises rapidly as the transfer function
decay time-scale is increased from zero to $\sim$1 hour. The
largest peak cross correlation function actually occurs with a decay
time-scale of  ~1.7 orbits (2 hours 45 mins) but the value of the
peak does not change greatly between decay values of $\sim$1 hour
and a few hours.

\subsection{Very short time-scales}
\label{sec:veryshort}

\begin{figure}
 \includegraphics[width=85mm,angle=0]{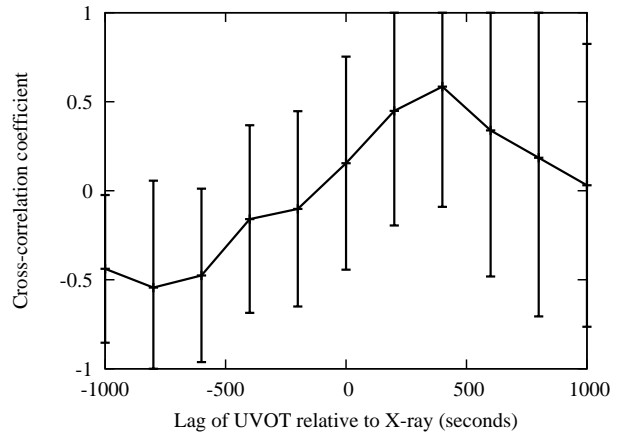}
\caption{The very short time-scale ($<1000$\,seconds) cross-correlation
 function between the X-ray and $uvw2$ bands calculated for the 2011
 August ToO observations.  The CCF shows a broad peak centred at a 400
 s lag of the $uvw2$ behind the X-ray.  The points are binned on 200s
 time-scales as in the lightcurve in Fig. \ref{fig:veryshortlc}.}
 \label{fig:vshortccf}
\end{figure}

In Fig. \ref{fig:vshortccf} we present the combined DCF (i.e. the weighted average of the DCFs for each segment) for the very short time-scale lightcurves
shown in Fig. \ref{fig:veryshortlc}. These data hint at a lag of the $uvw2$ band
by around 400s, but are not sufficient for any definite conclusion to
be drawn. They are, however, consistent with the DCF derived from the
2006 short time-scale monitoring (Fig. \ref{fig:ccfearly}) and with the
Chandra and HST monitoring of \citep{O'neill2006}.

\section{Accretion Disc Modelling}
\label{sec:AccDiscMod}
\begin{figure}
 \includegraphics[width=85mm,angle=0]{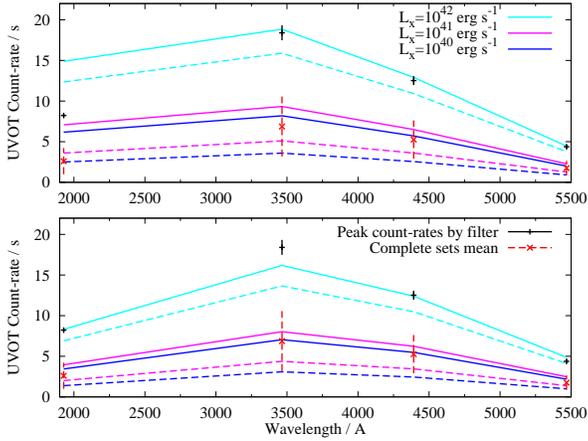}
 \caption{Spectral energy distributions: Top: The pure blackbody disc models plotted against the observed SEDs.  Bottom: The reddened blackbody disc model plotted against the observed SEDs.  Solid lines correspond to $\mathrm{\dot{m}}=10^{-3.0}\,\mathrm{\dot{m}}_{Edd}$ and the dashed lines to $\mathrm{\dot{m}}=10^{-3.5}\,\mathrm{\dot{m}}_{Edd}$.  All countrates within each plot are scaled by the same factor.}
 \label{fig:sed}
\end{figure}

As we have observations in 4 UVOT filters for each of the monitoring
observations in 2008/9 we are able to determine crude spectral
energy distributions (SEDs) which we show in Fig. \ref{fig:sed}. Here we show both an SED from the average of all of the data and an SED representative of the highest flux levels.  We can compare these SEDs with the prediction from a simple optically thick, geometrically thin, \citet{Shakura1973} accretion disc with constant accretion rate.

In our implementation we split the disc into 50000 logarithmically spaced segments between 6 R$_g$ and 60000 R$_g$, covering almost all of the emission from the far UV to far IR. We assume a black hole mass of $3.6 \times 10^{5}$ M$_\odot$ \citep{Peterson2005}, an accretion rate of 0.1 per cent \citep{Peterson2005}, a distance of 4.2 Mpc \citep{Thim2004} and place an X-ray point source at a height of $\sim$20 R$_g$ above the central black hole.  The albedo of the disc is assumed to be of order 0.3 \citep{Gierlinski2009} and altering this value would just result in a scaling of the input L$_x$.  We determine the integrated sum of the black body emission from each segment and fold it through the UVOT filter responses to determine the predicted count rates.

We see that although the predicted and observed count rates can agree reasonably well at the red end of the spectrum (Fig. \ref{fig:sed}), the observed $uvw2$ count rate is a factor $5$ below the expected count rate. Although we have tried very hard to remove the background accurately, it is possible that a small systematic error exists which could account for any differences at the red end of the spectrum. However, it is not possible to account for the difference at the blue end by such errors. There are two simple explanations for this, the disc is truncated at a larger radius than 6 R$_g$ or the observed spectrum is reddened by passage through some absorbing material. Given previous measurements of $A_V \sim$0.4 mag \citep{Lira1999} we prefer the reddening explanation. We require modest absorption of E($B-V$) $\sim$0.14 mag to reconcile the observed and predicted fluxes. For a standard Galactic dust to gas ratio this reddening corresponds to N$_H \sim1.0\times10^{21}$ cm$^{-2}$, which is quite consistent with the X-ray observations (Dwelly et al, in prep). Also \citet{Lira1999} note that the Balmer decrement derived from the ratio of the fluxes of the narrow components of $H_{\alpha}$ and $H_{\beta}$ is equivalent to an $A_{V} \sim$0.4 mag, which is consistent with our observation with an $A_{V} \sim$0.44 mag. The predicted countrates for a reddened face on disc are similar to the observed count rates, given the uncertainties in, eg, distance or disc inclination angle.

\section{Discussion}
\label{sec:Discussion}

We first list our main observational conclusions.
\begin{enumerate}
 \item
During monitoring from April 2008 to March 2009 with {\em Swift}, a
large increase in flux was detected from NGC\,4395 in the X-ray, UV and
optical wavebands. This high flux period lasted for over 60 days, reaching
X-ray flux levels at least 5 times greater than the previous average
levels, before returning to those previous levels.
 \item There is a strong correlation between the UV/optical variations
 with peak CCF values of $>0.9$, and all show similar levels of fractional variability.
  \item The variations between the hard band X-rays (2--10 keV) and all UVOT
 bands are highly correlated with a lag consistent with zero
 days. The strength of these correlations decreases with
 increasing UV/optical wavelength.
\item NGC\,4395 was also observed once per {\em Swift} orbit, every orbit, for 5
   days in 2006, using only the $b$ UVOT filter. The lag was again
   consistent with zero but the CCF was slightly asymmetric in the
   sense of a $b$-band lag of less than half a {\em Swift} orbit (i.e. $<$48mins).
\item By splitting individual {\em Swift} TOO observations into 200s bins, we find a weak correlation between the $uvw2$ and the X-ray lightcurves with a lag of $\sim$400 s.
\item The UVOT colours are consistent with the expectation from a standard optically thick accretion disc model, but only with the addition of a small amount of reddening.

\end{enumerate}

These {\em Swift} observations, together with previous less extensive
observations involving Chandra \citep{Desroches2006}, all indicate that the UV/optical
emission in NGC\,4395 probably lags behind the X-ray emission by a very
short time-scale, certainly less than 48 min and more likely closer to a
few hundred seconds. There are no observations which suggest that the
UV/optical emission as observed on short (i.e. hours-day) time-scales
leads the X-ray emission. A UV/optical lag of a few hundred seconds is
consistent with the light travel time to the UV/optical emission
region in the accretion disc and thus strongly suggests that the short
time-scale UV/optical variations arise from reprocessing of X-ray
emission.

If the UV/optical variability is driven mainly by reprocessing of
X-ray emission, it is necessary that the variable component of the
luminosity in the X-ray band exceeds that in the UV/optical bands. It
is hard to measure the relative luminosities precisely as our
observations do not cover all possible wavebands, but we can make an
approximate calculation. For example in Fig. \ref{fig:longlc} we see that the
UV/optical variability is greatest in the UVW2 band where the range of
fluxes is $\sim4 \times 10^{-15}$ ergs cm$^{-2}$ s$^{-1}$
\AA$^{-1}$. In order to obtain the total variable UV/optical flux
which might be driven by the X-rays we need to integrate over all
UV/optical bands. If we take a total bandwidth of 1000 \AA, and a flat
spectrum, we would obtain a variable flux of $\sim4 \times 10^{-12}$
ergs cm$^{-2}$ s$^{-1}$. For the Swift XRT a 2-10 keV flux of $1
\times 10^{-11}$ ergs cm$^{-2}$ s$^{-1}$, assuming a typical NGC\,4395
photon index of 1.6, corresponds to $\sim0.1$ counts s$^{-1}$. Thus
the variable 2-10 keV flux is $\sim4 \times 10^{-11}$ ergs cm$^{-2}$
s$^{-1}$. Even allowing for the fact that probably less than half of
the observed X-ray luminosity impinges upon the disc, this crude
estimate indicates that there is probably sufficient luminosity in the
X-ray variations to drive the UV/optical variability. 

One might argue that simply extending the UVW2 variability
over 1000 \AA \ underestimates the total variable UV/optical
luminosity. On the other hand the 2-10 keV flux is less, by factors of
a few, than the total irradiating X-ray flux, which may compensate.
Other methods of estimating the total variable luminosity in the
relevant bands, eg by modelling of the accretion disc, reach broadly
the same conclusion. We therefore conclude that although there is
probably not a great deal of room to spare, the luminosity in the
X-ray band is just about sufficient to drive the UV/optical
variability.

We note also that our treatment is rather simplistic. In reality the
enhanced UV/optical emission which is produced by X-ray irradiation
will probably lead to an enhanced flux of seed photons into the X-ray
emitting corona, and hence to further X-ray emission. Thus variations
in both bands may be prolonged in this feedback process. However as long as
the feedback is not too strong, the X-ray/UV-optical lag will remain
as a valid indicator of the approximate separation of the emission
regions.

The reason for the large outburst in 2008 is not entirely clear. \citet{Janiuk2011} suggest that disc radiation pressure instabilities, which might produce outbursts not too different from that seen, might occur in
AGN such as NGC\,4395 with a few year time-scale. \citet{Janiuk2011} suggest that an accretion rate greater than 0.025 Eddington, which is probably just above that of NGC\,4395, is required for the instability to be relevant in AGN. However given the uncertainties both in theoretical modelling and in estimating masses and accretion rates, this instability might still be relevant in NGC\,4395. Alternatively the apparent outburst may just be part of the normal stochastic variability of the AGN. Although the X-ray flux in the 2011 monitoring observations did not reach the same level as in 2008, the mean 2011 flux level is a good deal higher than in the second part of the 2008--2009 monitoring observations and there are other
similarities, e.g. softening of the spectrum with increasing flux (Dwelly et al, in prep). We discuss the detailed temporal variability in more detail elsewhere (M$^{\mathrm{c}}$Hardy et al, in prep).

\section*{Acknowledgments}

DTC thanks the School of Physics and Astronomy, Southampton University, for the award of a Mayflower Studentship. We also thank STFC for the award of a rolling grant which supports TD and for the award of an Advanced Fellowship which supports PU. PA acknowledges support from
Fondecyt 1110049. We thank Chris Done for useful discussions.

\bibliographystyle{mn_antonysmith}
\bibliography{references}

\end{document}